\documentclass[10pt]{article}
\usepackage{amssymb}
\title{A First-Principles Implementation of Scale Invariance Using Best Matching}
\author{Hans Westman\thanks{\tt Email: hwestman@physics.usyd.edu.au}\\[2mm]%
{\small \it Physics Building A28, University of Sydney, NSW 2006}\\%
{\small \it Sydney, New South Wales, Australia}\\[2mm]%
}%
\date{{\small \today }}%
\begin{document}
\maketitle
\abstract{We present a first-principles implementation of {\em spatial} scale invariance as a local gauge symmetry in geometry dynamics using the method of best matching \cite{BB82}. In addition to the 3-metric, the proposed scale invariant theory also contains a 3-vector potential $A_k$ as a dynamical variable. Although some of the mathematics is similar to Weyl's ingenious but physically questionable theory \cite{Weyl1918,Weyl1922,Bach1921}, the equations of motion of this new theory are second order in time-derivatives. Thereby we avoid the problems associated with fourth order time derivatives that plague Weyl's original theory. It is tempting to try to interpret the vector potential $A_k$ as the electromagnetic field. We exhibit four independent reasons for not giving into this temptation. A more likely possibility is that it can play the role of ``dark matter''. Indeed, as noted in \cite{monda,mondb} scale invariance seems to play a role in the MOND phenomenology. Spatial boundary conditions are derived from the free-endpoint variation method and a preliminary analysis of the constraints and their propagation in the Hamiltonian formulation is presented.}
\section{Introduction}
In practice we are never able to measure an objects ``intrinsic'' length. Rather, a measurement of the length of some object is a demonstration that $x$ reference rulers can be fitted alongside it. But if both the reference ruler and the physical object are simultaneously rescaled, the measurement outcome would remain unaltered. Thus, we cannot measure an objects ``intrinsic'' length but only the ratio of lengths. Therefore, on epistemological grounds, there seems to be good reasons to expect a fundamental theory to be scale invariant so that the fundamental equations of motion make no reference to intrinsic sizes. Furthermore, since length measurements are ultimately done locally in space, it also seems reasonable to expect a fundamental theory to be also invariant with respect to {\em local} scale transformations. 

This heuristic argument, based on basic epistemology, is very similar to the well-known arguments about absolute space. In practice we are never able to measure a body's ``absolute'' position. Instead we can only measure a body's position relative to other bodies. Therefore, on epistemological grounds we expect that a fundamental theory should not make reference to absolute positions. This heuristic reasoning, dating back at least to the famous Newton-Leibniz debates over absolute space, was undoubtedly important for the discovery of general relativity in which absolute positions play no role.

Given that this type of heuristic reasoning has proven very successful, it is only natural to try to continue this process of ``epistemological refinement'' of physical theories. On this view, the fact that general relativity is not scale invariant can be considered an ``epistemological defect''. In fact, a number of scale invariant theories has indeed been put forth in the literature. An important early attempt in 1918 is by Hermann Weyl \cite{Weyl1918,Weyl1922,Livingrevs} and more recently by Mannheim \cite{Mannheim}. Furthermore, Barbour has constructed scale invariant particle model whose predictions  seem to be approximately that of Newtonian gravity \cite{Barbour02partmod}.

In \cite{Barbour04a,Barbour04b} Anderson {\em et. al.} presented an intriguing first-principles derivation of York's conformal technique \cite{Yorka,Yorkb,Yorkc} for solving the Hamiltonian constraint in general relativity by enforcing {\em volume preserving} spatial conformal transformations as a gauge symmetry. They start from a special form of the Einstein-Hilbert action called the Baierlein-Sharp-Wheeler action \cite{BSW62}. Then they perform an arbitrary volume preserving conformal transformation. Since the Beierlein-Sharp-Wheeler action is not gauge invariant with respect to conformal transformations both the scale parameter $\phi$ and its velocity $\dot{\phi}$ will appear in the so obtained conformalized action. From the free-endpoint variation method they argue that $\dot{\phi}$ and $\phi$ should be varied independently. As a intriguing consequence, the constant mean curvature foliation time gauge as well as the Lichnerowicz-York equation falls out. 

Although this work sheds light on the importance of conformal structures in geometry dynamics for solving the Hamiltonian constraint, this trick of introducing conformal invariance by conformalizing the Beierlein-Sharp-Wheeler action is analogous to the way one can turn the Klein-Gordon equation in flat spacetime into a 4-diffeomorphism invariant theory by parametrizing the Minkowski coordinates \cite{Kiefer,SonegoWestman}. One of the purposes of this this paper is to investigate whether conformal invariance can be implemented from first principles and not by starting from the conformalized Baierlein-Sharp-Wheeler action. Of course, this is a more risky project and there is no guarantee that the theory presented here is empirically adequate. 
\section{Mathematical implementation of scale invariance}
We shall now start building in scale invariance in our theory from first principles. Consider then a ruler. If we move the ruler from a point $A$ to $B$, how can we be sure that the ruler has the same length after it has been moved? How do we know that there is not a new kind of force that rescales the ruler when we move it from point to point is space? If the rescaling effect is a universal so that all objects, irrespective of internal composition, are rescale in the same way, then it is pointless to introduce another reference ruler to check to see if the length of the first ruler has changed. This is so since by assumption the new field will have rescaled both rulers in the same way. Thus, the rescaling effect will be locally unobservable.\footnote{Of course, there might be interesting global effects. We will shortly discuss this in section \ref{hydrogen}.} 

Note however that a measurement of an angle reduces to a measurement of ratios of lengths of the sides of a triangle. Angles are therefore already fully relational quantities (as is also indicated by their lack of dimension) and can therefore be unambiguously measured locally. Thus, angles have a different epistemological status than lengths.

The key mathematical idea is therefore the following: allow for a connection $\Gamma^k_{\ ij}$ that preserves angles between parallely transported vectors (representing ideal rulers) but not necessarily their individual lengths. The Levi-Civita connection is too restrictive for this purpose since it preserves both lengths and angles. Instead we should only require that 
\begin{eqnarray}
\frac{D}{Ds}\left(\frac{U^iV^jg_{ij}}{|U||V|}\right)=X^k\nabla_k\left(\frac{U^iV^jg_{ij}}{|U||V|}\right)=0
\end{eqnarray}
for all $X^k$, and all parallely transported $U^i$ and $V^i$, i.e. $\frac{D}{Ds}U^i=\frac{D}{Ds}V^i=0$, where $|U|=\sqrt{U^iU^jg_{ij}}$ and $|V|=\sqrt{V^iV^jg_{ij}}$. By expanding this expression, rearranging terms, and making use of that the vectors are parallely transported we get
\begin{eqnarray}
\nabla_kg_{ij}\left[\frac{U^iV^j}{U\cdot V}-\frac{1}{2}\frac{U^iU^j}{|U|^2}-\frac{1}{2}\frac{V^iV^j}{|V|^2}\right]=0
\end{eqnarray}
The only way to satisfy this equation is if
\begin{eqnarray}
\nabla_kg_{ij}=A_kg_{ij}.
\end{eqnarray}
Thus, the connection will in general be non-metrical, i.e. $\nabla_kg_{ij}\neq 0$. If we assume vanishing torsion $\Gamma^k_{\ ij}=\Gamma^k_{\ ji}$ and make use of that $\nabla_kg_{ij}=\partial_kg_{ij}-\Gamma^l_{\ ki}g_{lj}-\Gamma^l_{\ kj}g_{il}$ this implies that the connection takes the form
\begin{eqnarray}
\Gamma^{k}_{\ ij}=\left\{\begin{array}{c}k\\ij\end{array}\right\}-\frac{1}{2}(\delta^k_iA_j+\delta^k_jA_i-g_{ij}g^{kl}A_l)
\end{eqnarray}
where $\left\{\begin{array}{c}k\\ij\end{array}\right\}$ is the Levi-Civita connection. Under a conformal transformation $g_{ij}\rightarrow e^{\theta}g_{ij}$ the Levi-Civita connection transforms in the following way:
\begin{eqnarray}
\left\{\begin{array}{c}k\\ij\end{array}\right\}\rightarrow\left\{\begin{array}{c}k\\ij\end{array}\right\}+\frac{1}{2}(\delta^k_i\partial_j\theta+\delta^k_j\partial_i\theta-g_{ij}g^{kl}\partial_l\theta)
\end{eqnarray}
Thus, if the vector potential transforms as $A_i\rightarrow A_i+\partial_i\theta$, then the full non-metrical connection $\Gamma^k_{\ ij}=\left\{\begin{array}{c}k\\ij\end{array}\right\}-\frac{1}{2}(\delta^k_iA_j+\delta^k_jA_i-g_{ij}g^{kl}A_l)$ is not only conformally covariant but also conformally {\em invariant} since all the $\theta$'s cancel out. Our connection is therefore unchanged by a conformal transformation. This, in turn, means that the corresponding Riemann and Ricci curvatures 
\begin{eqnarray}
R^k_{\ lij}&=&\partial_i\Gamma^k_{\ jl}-\partial_j\Gamma^k_{\ il}+\Gamma^k_{\ in}\Gamma^n_{\ jl}-\Gamma^k_{\ jn}\Gamma^n_{\ il}\\
R_{lj}&=&R^k_{\ lkj}
\end{eqnarray}
are also conformally invariant. However, $R=g^{ij}R_{ij}$ is not invariant since it involves contraction with the metric which comes with conformal weight of $-1$, i.e. $R\rightarrow e^{-\theta}R$. Note also that $R_{ij}\neq R_{ji}$ in general. This is so since we are not dealing with a metrical connection.

The reader familiar with Weyl's 1918 theory \cite{Weyl1918,Weyl1922} will see that, up till now, the mathematics is identical to Weyl's theory. One important exception is that we are implementing {\em spatial} scale invariance $g_{ij}\rightarrow e^\theta g_{ij}$ while Weyl implemented spatiotemporal scale invariance $g_{\mu\nu}\rightarrow e^\theta g_{\mu\nu}$. As we shall see, this mathematical difference is going to be crucial when we try to construct a scale invariant dynamics of the theory and allows us to obtain a theory which has field equations of second order in time. 
\section{Einstein objections to Weyl's theory}
Einstein's reaction to Weyl's 1918 unified theory was mixed \cite{Livingrevs}. On the one hand side Einstein was genuinely impressed by its mathematical beauty and ingenuity calling it ``a stroke of genius of first rank''. But on the other hand Einstein expressed doubts about the theory as a physical theory. In this section we are going to review Einstein's main misgivings about Weyl's 1918 theory.

\subsection{Fourth order field equations}
First Einstein remarked that the ``gravitational field equations will be of fourth order, against which speaks all experience until now''. More specifically, fourth order theories in general imply a Hamiltonian that does not have a lower bound \cite{pagini87}. This gives rise to ``instability'' problems which is not desirable for a physical theory. In contrast to Weyl we shall only implement spatial scale invariance and in the theory we shall present below, the field equations are guaranteed to be second order in time-derivatives thereby avoiding the problems associated with higher-order derivatives. 
\subsection{Proper time}\label{proptime}
Einstein also raised an objection concerning proper time. His objection runs as follows. Assume that the line element $d\tau^2=g_{\mu\nu}dx^\mu dx^\nu$ corresponds to the proper time of a clock. Since proper time (as read off from a ideal clock) is something observable it should have a unique predictable value in a theory. Secondly, assume with Weyl that conformally related metrics are physically equivalent, i.e. $g_{\mu\nu}$ and $e^\theta g_{\mu\nu}$ describe the same physical situation. However, the line element also gets transformed $d\tau^2\rightarrow e^\theta d\tau^2$ and as a consequence the theory cannot make a unique prediction regarding what proper time a clock will show when moved along some path in spacetime. Therefore, there is a flaw in Weyl's theory.

One obvious way to avoid such a conclusion is to challenge the assumption that proper time of a physical clock can be obtained from the line element of the metric. Instead Weyl proposed that the proper time should be $d\tau^2=Rg_{\mu\nu}dx^\mu dx^\nu$, where $R$ is the scalar curvature in Weyl's theory. The object $Rg_{\mu\nu}dx^\mu dx^\nu$ has conformal weight zero and consequently does not transform under conformal transformations and in this way a unique proper time is obtained. However, this choice is hard to motivate on physical grounds. What if $R\approx0$ for example? 

In another attempt to avoid Einstein's objection Weyl also maintained that before anything definite can be said about proper time, a theory of clocks needs to be developed {\em within} the theory. However, to the best of our knowledge, Weyl never developed such a theory of clocks and Einstein's objection was left unresolved.

Nevertheless, Einstein's objection concerning proper time might not be as serious as first thought. We are not aware of any later attempts to develop a theory of clocks within Weyl's theory so here we shall briefly sketch such a theory. The simplest theoretical clock one can imagine is the lightclock.\footnote{I am grateful to Lucian Hardy for suggesting this as a possible resolution to Einstein's ``proper time'' objection.} Consider then any spacetime trajectory $\gamma$. Closely around it we put mirrors that would make a light-beam bounce back and forth crossing the trajectory many times. The proper time measured by this light clock would be the number of times the light-ray crosses over the trajectory $\gamma$.

However, we need to make sure that the mirrors are positioned so that they remain at the same proper distance throughout the trajectory. This creates an immediate problem: how do we define an notion of equidistance in a conformally invariant theory in which distance is but a gauge degree of freedom? This problem is only apparent though. Weyl's theory to contains the necessary mathematical machinery: At the beginning of the trajectory we choose an arbitrary (small) vector representing a reference ruler. Then we Fermi-Walker transport\footnote{A parallel transport of the vector would not be appropriate since in general the vector will not remain on the spatial slice orthogonal to the worldline of the lightclock.} this reference ruler along the curve using the Weyl non-metrical connection. In this way we can define a notion of equidistance within Weyl's theory and the notion of proper time becomes well-defined and presumably gauge-independent.

For our theory to be viable it also needs to circumvent Einstein's objection regarding proper time. However, since a similar analysis of lightclocks would require a universal lightcone structure, something which has yet to be demonstrated for the our theory, we postpone a full analysis to a future paper.

\subsection{Hydrogen spectral lines}\label{hydrogen}
The connection $A_\mu$ in Weyl's theory might be non-trivial $F_{\mu\nu}=\partial_\mu A_\nu-\partial_\nu A_\mu\neq0$. This has the following implication \cite{Livingrevs}: If we move two a hydrogen atoms from a point $A$ to B in spacetime, but along different paths, then the final size of the hydrogen atoms might differ. As a concequence the spectral lines would be shifted. However, experiments provide very tight bounds on deviations from the standard predictions. We are not aware of any attempts of a numerical quantification of this effect but it seems plausible that, if the vector potential $A_\mu$ is taken to represent the electromagnetic field, then current experiments would probably rule out the theory. 

However, as is argued in section \ref{Discussion}, the vector potential $A_\mu$ cannot represent the electromagnetic field but could instead be thought of as a candidate for ``dark matter''. But if $A_\mu$ is not the electromagnetic field, its typical strength is not known and could very well be very small. It would therefore be worth speculating about whether a Weyl-type theory could explain the (controversial) claim of a tiny increase of the fine structure constant over the last 10 billion years. The fine structure constant is estimated by observations of spectral lines from distant astronomical objects and a varying fine structure constant could perhaps instead be reinterpreted as a Weyl-type rescaling effect.
\section{Best matching}
For Weyl the task to find a scale and diffeomorphism invariant action for his theory was rather straightforward: you need a spacetime scalar density with conformal weight zero. This would ensure 4-diffeomorphism invariance as well as conformal invariance as gauge symmetries. However, our approach is explicitly 3-dimensional in nature and therefore we have to resort to a different strategy to construct an action. Such a strategy, called ``best matching'', has been developed by Julian Barbour and collaborators (see e.g. \cite{BB82,Barbour02partmod}). The technique of ``best matching'' yields specific Lagrangians and bosonic gauge field theories (e.g. Maxwell's, Yang-Mill's theories, and general relativity) are examples of ``best matched'' theories \cite{Anderson}.\footnote{As of date, it has not been shown that spinors fields are compatible with the principle of best-matching. The main difficulty is the following: Actions for four-component Dirac spinor fields are not quadratic but linear in the field velocities while best-matched actions are always quadratic. A possible way to proceed could be to consider van der Waerden's reformulation of the Dirac field in which the equations for the two-component spinor is second order in spatio-temporal derivatives.}

When carrying out the best-matching procedure in general relativity (and most importantly requiring a ``local square-root'' \cite{BarbourRWR}) we end up with the following Lagrangian
\begin{eqnarray}
L=\int d^3x\sqrt{g}\sqrt{T V}=\int d^3x\sqrt{g}\sqrt{G^{ijkl}\left(\dot{g}_{ij}-\mathcal{L}_{\overrightarrow{N}}g_{ij}\right)\left(\dot{g}_{kl}-\mathcal{L}_{\overrightarrow{N}}g_{kl}\right)R}
\end{eqnarray}
where $T=G^{ijkl}\left(\dot{g}_{ij}-\mathcal{L}_{\overrightarrow{N}}g_{ij}\right)\left(\dot{g}_{kl}-\mathcal{L}_{\overrightarrow{N}}g_{kl}\right)$, $G^{ijkl}=g^{ik}g^{jl}-g^{ij}g^{kl}$, and $V=R$. This Lagrangian, first discovered by Baierlein, Sharp, and Wheeler \cite{BSW62}, and later significantly illuminated by Barbour {\em et. al.} \cite{BarbourRWR}, yields the same equations of motion as the Einstein-Hilbert action. It has a peculiar local square root structure, i.e. the square root is taken before the integration over space. This is something unique to general relativity \cite{BarbourRWR} and is shared neither by the Barbour-Bertotti model \cite{BB82}, nor Yang-Mill's theories such as Maxwell's theory even when written in the mathematically equivalent best-matched form \cite{BB82}. However, this peculiar local square root leads to foliation invariance as a unexpected additional symmetry for a narrow set of potentials \cite{BarbourRWR}. In this way 4-diffeomorphism invariance is retained. The emergence of a universal lightcone structure is also crucially dependent on having a local square root form \cite{BarbourRWR}.

As we shall see in section \ref{localsqr}, although the the local square root structure in general relativity is somewhat mysterious, it seems to be very natural when local spatial scale invariance is demanded.
\subsection{Kinetic term}
We shall assume that the Lagrangian density for our theory has a similar square root form $\mathcal{L}=\sqrt{g}\sqrt{T V}$. We can now immediately write down the kinetic terms
\begin{eqnarray}
T_g&=&G^{ijkl}\left(\dot{g}_{ij}-\mathcal{L}_{\overrightarrow{N}}g_{ij}-\phi g_{ij}\right)\left(\dot{g}_{kl}-\mathcal{L}_{\overrightarrow{N}}g_{kl}-\phi g_{kl}\right)\\
T_A&=&g^{ij}\left(\dot{A}_i-\mathcal{L}_{\overrightarrow{N}}A_i-\partial_i\phi\right)\left(\dot{A}_j-\mathcal{L}_{\overrightarrow{N}}A_j-\partial_j\phi\right)
\end{eqnarray}
A new auxiliary field $\phi$ has been added in order to impose the best-matching condition (obtained by varying the Lagrangian with respect to $\phi$) corresponding to the new conformal gauge symmetry. Note that these are basically the only natural choices available for quadratic kinetic terms.

These kinetic terms are conformally covariant only if $\phi\rightarrow\phi+\dot{\theta}-N^k\partial_k\theta$ and $N^k\rightarrow N^k$ under a conformal transformation $g_{ij}\rightarrow e^{\theta}g_{ij}$ and $A_i\rightarrow A_i+\partial_i\theta$. Given this transformation rule of the auxiliary field $\phi$ the kinetic terms transforms according to 
\begin{eqnarray}
T_g&\rightarrow&T_g\\
T_A&\rightarrow&e^{-\theta}T_A.
\end{eqnarray}
Thus, $T_g$ and $T_A$ has conformal weight $0$ and $-1$ respectively. This means that if we want to ensure conformal invariance, we cannot simply add these kinetic terms together as one would normally do (see e.g. \cite{BarbourRWR}). Instead, we need to multiply one of the kinetic terms with a scalar quantity with the appropriate scalar weight before adding them. The only natural candidate seems to be the scalar curvature $R$. Thus, the most natural kinetic term whould have the form
\begin{eqnarray}
T=T_gR+aT_A
\end{eqnarray}
where a is an arbitrary constant. The kinetic term will then have conformal weight $-1$.
\subsection{Potential}
In order to ensure conformal invariance we need to make sure that the action has conformal weight zero. $\sqrt{g}$ has weight $+3/2$ and this needs to be compensated by $\sqrt{TV}$. Since the kinetic term $T$ has conformal weight $-1$ we deduce that the potential $V$ must have conformal weight $-2$. 

There are only a few natural candidates for 3-scalars that have the right conformal weight \cite{Livingrevs}:\footnote{We do not consider higher order curvature scalars which involves higher order spatial derivatives.}
\begin{eqnarray}
R^2\qquad R_{ij}R^{ij}\qquad F_{ij}F^{ij}\qquad R_{ij}F^{ij}\qquad R^k_{\ lij}R_k^{\ lij}
\end{eqnarray}
where $F_{ij}=\partial_iA_j-\partial_jA_i$. We will keep the discussion general here and take the potential $V$ to be a linear combination of all those terms.
\begin{eqnarray}
V=bR^2+cR_{ij}R^{ij}+dF_{ij}F^{ij}+eR_{ij}F^{ij}+fR^k_{\ lij}R_k^{\ lij}
\end{eqnarray}
Note that $R_{ij}$ is not necessarily a symmetric tensor and therefore $R_{ij}F^{ij}$ is not identically zero. This is because we are not dealing anymore with the Levi-Civita metrical connection. 

The action for the theory is then
\begin{eqnarray}
\mathcal{L}=\sqrt{g}\sqrt{(T_gR+aT_A)V}.
\end{eqnarray}
It is immediately recognized that, since only first order time-derivatives enter in the Lagrangian density, the field equations will be second order in time. Thus, we have avoided the ``instability'' problems connected to higher-order time-derivatives.

\subsection{Fixing numerical values of the constants}\label{constantfix}
One way to try to determine the constants $a,b,c,d,e,f$ would be to require foliation invariance (i.e. that the Hamiltonian constraint will propagate). It was noted in \cite{BarbourRWR} that the requirement of foliation invariance (i.e. the propagation of the Hamiltonian constraint) quite uniquely picks out the Baierlein-Sharp-Wheeler action. We hope that something similar will happen for our conformally invariant theory. At this point it is not clear if it is possible to make the Hamiltonian constraint propagate. We will postpone a full analysis for a later paper.

\subsection{Local square-roots and local scale invariance}\label{localsqr}
The best-matched Lagrangian for the Maxwell field looks like \cite{BB82}
\begin{eqnarray}
L=\sqrt{\int d^3x\delta^{ij}(\dot{A}_i-\partial_i\phi)(\dot{A}_j-\partial_j\phi)\left(E-\int d^3x \delta^{ij}B_iB_j\right)}
\end{eqnarray}
where $A_i$ is the electromagnetic vector potential, $B_i=\epsilon_{ijk}\partial_jA_k$ the magnetic field, $E$ is the total energy, and $\phi=A_0$. Notice that the square-root is taken after we have integrated over space. The {\em local} square-root structure in the Beierlein-Sharp-Wheeler action, where the square-root is taken {\em before} the integration, is therefore quite surprising. It accounts for the emergence of spacetime diffeomorphism invariance and a universal lightcone structure but at the same time it also signifies an important structural difference from other best-matched theories, e.g. the Maxwell field \cite{BB82} which has a global square root-structure as just mentioned. It would therefore be interesting to see if there is some fundamental reason form the local square-root in general relativity.

Indeed, it seems that local scale invariance can shed some light on this issue: A global square-root structure is rather unnatural in the scale invariant framework. A confrmally invariant Lagrangian with a global square-root could look like

\begin{eqnarray}
L=\sqrt{\int d^3x\sqrt{g}(T_gR+aT_A)V_1\left(E-\int d^3x\sqrt{g}V_2\right)}
\end{eqnarray}
where $V_1$ and $V_2$ are potentials. They must have conformal weight $-1/2$ and $-3/2$ respectively. The only such scalars that can be formed include square-roots and absolute signs, e.g. $\sqrt{|R|}$ and $(F^2)^{1/4}$. These Lagrangians, although they cannot be excluded {\em \`a priori}, appear rather uggly and unnatural. Therefore it seems that if local spatial scale invariance is imposed, then a local square-root structure is natural.

\section{Machian boundary conditions from free-endpoint variations}
Let us now consider a variation of the action $S=\int d^4x\mathcal{L}$ with respect to the auxiliary fields $N^k$ and $\phi$. Making use of Gauss theorem yields
\begin{eqnarray}
\delta_\phi S&=&\int_V d^4x\left(\frac{\partial\mathcal{L}}{\partial\phi}-\partial_\mu\frac{\partial\mathcal{L}}{\partial \partial_\mu \phi}\right)\delta\phi+\int_{\partial V}dA n_\mu\frac{\partial\mathcal{L}}{\partial_\mu\phi}\delta\phi\\
\delta_{N^k} S&=&\int_V d^4x\left(\frac{\partial\mathcal{L}}{\partial N^k}-\partial_\mu\frac{\partial\mathcal{L}}{\partial \partial_\mu N^k}\right)\delta N^k+\int_{\partial V}dA n_\mu\frac{\partial\mathcal{L}}{\partial_\mu N^k}\delta N^k
\end{eqnarray}
where $V$ stands for a four-dimensional region in spacetime, $\partial V$ its three-dimensional boundary,  and $n_\mu$ the corresponding unit normal vector. The variational principle requires that the variation $\delta S$ is zero for all for all variations $\delta \phi$ and $\delta N^k$. Normally one makes the restriction that $\delta\phi=\delta N^k=0$ on the boundary. However, since we are dealing with unphysical auxiliary fields $N^k$and $\phi$ whose only purpose is to enforce the ``best matching'' condition, this restriction is not compelling. Indeed, to enforce an arbitrary restriction on the variation of the auxiliary fields on the boundary amounts to introducing an arbitrary absolute element in the theory, something which should be avoided in a relational theory. Instead one should allow the variation $\delta\phi$ to be arbitrary even on the boundary \cite{Barbour02partmod}. This is called the free-endpoint variation method\footnote{Free endpoint variations are encountered in numerous problems in analytical mechanics where the endpoints cannot be fixed {\em \`a priori}. A standard example is a flexible hanging beam whose position and orientation are fixed at one end but free in the other. The action is the total energy which consists of potential energy and ``bending'' energy. The variation at the non-fixed end is unknown {\em \`a priori} and must be allowed to be completely free. The shape of the beam is found by minimizing the total energy. This implies non-trivial boundary conditions for the non-fixed end.} and yields the following equations of motion
\begin{eqnarray}
\frac{\partial\mathcal{L}}{\partial\phi}-\partial_\mu\frac{\partial\mathcal{L}}{\partial \partial_\mu \phi}&=&0\qquad\left. n_\mu\frac{\partial\mathcal{L}}{\partial_\mu\phi}\right|_{x\in\partial V}=0\\
\frac{\partial\mathcal{L}}{\partial N^k}-\partial_\mu\frac{\partial\mathcal{L}}{\partial \partial_\mu N^k}&=&0\qquad\left. n_\mu\frac{\partial\mathcal{L}}{\partial_\mu N^k}\right|_{x\in\partial V}=0.
\end{eqnarray}
If, as is usually done, the boundary $\partial V$ is chosen so that $n_\mu=\delta^0_\mu$ for $\lambda=\lambda_1$, $n_\mu=-\delta^0_\mu$ for $\lambda_2$ and $n_\mu=(0,n_i)$ for $\lambda_1<\lambda<\lambda_2$ then these equations become
\begin{eqnarray}
\frac{\partial\mathcal{L}}{\partial\phi}-\partial_i\frac{\partial\mathcal{L}}{\partial \partial_i \phi}&=&0\qquad\left. n_k\frac{\partial\mathcal{L}}{\partial_k\phi}\right|_{\begin{array}{c}x\in\partial V\\\lambda\in[\lambda_1 \lambda_2]\end{array}}=0\\
\frac{\partial\mathcal{L}}{\partial N^k}-\partial_i\frac{\partial\mathcal{L}}{\partial \partial_i N^k}&=&0\qquad\left. n_l\frac{\partial\mathcal{L}}{\partial_l N^k}\right|_{\begin{array}{c}x\in\partial V\\\lambda\in[\lambda_1 \lambda_2]\end{array}}=0
\end{eqnarray} 
where we have made use of the fact that the Lagrangian does not depend on the time-derivatives of the auxiliary fields so that $\frac{\partial\mathcal{L}}{\partial\dot{\phi}}\equiv0\equiv\frac{\partial\mathcal{L}}{\partial\dot{N}^k}$. Using the definition (see eqs. (\ref{gijmomenta})-(\ref{Akmomenta}) of section \ref{constraints}) of the canonical momenta we can rewrite the spatial boundary conditions as
\begin{eqnarray}
\left.\pi^kn_k\right|_{\begin{array}{c}x\in\partial V\\\lambda\in[\lambda_1 \lambda_2]\end{array}}&=&0\\
\left. \pi^i_{\ j}n_i\right|_{\begin{array}{c}x\in\partial V\\\lambda\in[\lambda_1 \lambda_2]\end{array}}&=&0
\end{eqnarray} 
Thus, the method of free endpoint variation yields important spatial boundary conditions for the fields $g_{ij}$ and $A_k$. The first constraint can be interpreted as there being zero ``charge'' in the universe (see equation (\ref{gauss})). The second constraint implies that the assymptotic ADM-type momentum and angular momentum must be zero for the whole universe, something which is expected in a Machian theory.\footnote{For explicit expressions of the ADM momenta in general relativity see \cite{Wald} and \cite{Poisson}.}

That {\em spatial} boundary conditions follow from the free endpoint variation method seems not to have been noted in the literature before. However, it should be noted that it is normally argued that only a spatially closed universe is compatible with Mach's principle. In such a case we must impose periodic spatial boundary conditions and therefore the issue of boundary conditions does not arise.
%
\section{Preliminary analysis of the constraints}\label{constraints}
The momenta conjugate to $g_{ij}$ and $A_i$ are
\begin{eqnarray}
\pi^{ij}&\equiv&\frac{\partial\mathcal{L}}{\partial\dot{g}_{ij}}=\sqrt{g}RG^{ijkl}(\dot{g}_{kl}-\mathcal{L}_{\overrightarrow N}g_{kl}-\phi g_{kl})\sqrt{\frac{V}{T_gR+aT_A}}\label{gijmomenta}\\
\pi^i&\equiv&\frac{\partial\mathcal{L}}{\partial\dot{A}_i}=\sqrt{g}ag^{ij}(\dot{A}_j-\mathcal{L}_{\overrightarrow N}A_j-\partial_j\phi)\sqrt{\frac{V}{T_gR+aT_A}}\label{Akmomenta}
\end{eqnarray}
Since the Lagrangian is independent of $\dot{N}^k$ and $\dot{\phi}$ their canonical momenta vanishes identically
\begin{eqnarray}
\pi^k_{\overrightarrow N}&\equiv&0\\
\pi_{\phi}&\equiv&0
\end{eqnarray}
Thus we have identified two primary constraints. We need these two constraints to propagate $\frac{d}{d\lambda}\pi^k_{\overrightarrow N}=0=\frac{d}{d\lambda}\pi_{\phi}$. By rewriting the Lagrangian equations of motion 
\begin{eqnarray}
0&=&\frac{\partial\mathcal{L}}{\partial\phi}-\partial_k\frac{\partial\mathcal{L}}{\partial \partial_k\phi}-\frac{d}{d\lambda}\frac{\partial\mathcal{L}}{\partial \dot{\phi}}=\frac{\partial\mathcal{L}}{\partial\phi}-\partial_k\frac{\partial\mathcal{L}}{\partial \partial_k\phi}-\frac{d\pi_\phi}{d\lambda}\\
0&=&\frac{\partial\mathcal{L}}{\partial N^l}-\partial_k\frac{\partial\mathcal{L}}{\partial \partial_kN^l}-\frac{d}{d\lambda}\frac{\partial\mathcal{L}}{\partial \dot{N}^l}=\frac{\partial\mathcal{L}}{\partial N^l}-\partial_k\frac{\partial\mathcal{L}}{\partial \partial_kN^l}-\frac{d\pi_{\overrightarrow{N}}}{d\lambda}
\end{eqnarray}
we see that the propagation of the constraints $\pi^k_{\overrightarrow N}$ and $\pi_{\phi}$ can be ensured if 
\begin{eqnarray}
\frac{d\pi_\phi}{d\lambda}&=&\frac{\partial\mathcal{L}}{\partial\phi}-\partial_k\frac{\partial\mathcal{L}}{\partial \partial_k\phi}=0\\
\frac{d\pi_{\overrightarrow{N}}}{d\lambda}&=&\frac{\partial\mathcal{L}}{\partial N^l}-\partial_k\frac{\partial\mathcal{L}}{\partial \partial_kN^l}=0.
\end{eqnarray}
After some calculation making use of the definitions for the canonical momenta and recalling that these are tensor densities rather than tensors it can be shown that these constraints take the form
\begin{eqnarray}
\pi^{ij}g_{ij}-\nabla_k\pi^k=0\label{gauss}\\
\nabla_i\pi^i_{\ j}+\frac{1}{2}\pi^iF_{ij}=0\label{diff}
\end{eqnarray}
These are, in Dirac's terminology, secondary constraints. Since the symmetries correspoding to these constraints (3-diffeomorphism invariance and spatial conformal invariance) is manifest in the Lagrangian it is clear that these constraints will propagate. According to the first constraint, which resembles Gauss law, the trace of $\pi^{ij}$ acts as a source for the ``electric'' field $\pi^i$. The second constraint is similar to the one in standard geometry dynamics but is now modified with an extra term which is familiar from quantum gravity in the Ashtekar variables (see e.g. \cite{Kiefer}).

Since we are dealing with a reparametrization invariant Lagrangian we can immediately read off the following quadratic primary constraint from the definitions of the canonical momenta:
\begin{eqnarray}
\frac{1}{\sqrt{g}}aG_{ijkl}\pi^{ij}\pi^{kl}+\frac{1}{\sqrt{g}}Rg_{ij}\pi^i\pi^j-a\sqrt{g}RV=0
\end{eqnarray}
where $G_{ijkl}G^{klmn}=\delta^m_i\delta^n_j$. This is the Hamiltonian constraint which arises from the global reparametrization invariance of the action\footnote{Whether or not the action is also locally reparametrization invariant, so that we would have a many fingered time as in general relativity, depends on whether the Hamiltonian constraint propagates or not.}. The total Hamiltonian thus looks like
\begin{eqnarray}
H=\int d^3x N\mathcal{H}+N^k\mathcal{H}_k+u\mathcal{C}
\end{eqnarray}
where 
\begin{eqnarray}
\mathcal{H}&=&\frac{1}{\sqrt{g}}aG_{ijkl}\pi^{ij}\pi^{kl}+\frac{1}{\sqrt{g}}Rg_{ij}\pi^i\pi^j-a\sqrt{g}RV\approx0\\
\mathcal{H}_j&=&\nabla_i\pi^i_{\ j}+\frac{1}{2}\pi^iF_{ij}\approx0\\
\mathcal{C}&=&\pi^{ij}g_{ij}-\partial_k\pi^k\approx0
\end{eqnarray}
represents the Hamiltonian, momentum, and conformal constraints respectively. $N$, $N^k$, and $u$ are Lagrangian multipliers.

It would be interesting to see whether the Hamiltonian constraint, with suitable numerical values for the constants $a,b,c,d,e$ and $f$, can be made to propagate. If the Hamiltonian constraint does not propagate and more secondary constraints must be introduced, the theory could turn out to be inconsistent (e.g. we end up with more constraints than unknowns or that $1=0$). Or perhaps the Hamiltonian constraint will perhaps not be first class. Or perhaps the constraints do close but not according to the Dirac-Teitelboim algebra so that a solution can not be embedded in a spacetime \cite{Teitelboim73}. These are very important issues but a proper analysis will unfortunately involve lengthy calculations and will be postponed to a forthcoming paper.

\section{Discussion}\label{Discussion}
Our scale invariant theory contains more degrees of freedom than conventional geometry dynamics. In addition to the gravitational field $g_{ij}$ it also contains the vector field $A_k$. For this theory to be empirically adequate it is imperative that the new degrees of freedom are actually represented in nature in some form or another. Given the mathematical similarity of the vector potential $A_k$ and the electromagnetic field it is tempting to try to identify $A_k$ with the electromagnetic field.\footnote{Indeed, Weyl considered his theory as a unification of the gravitational and electromagnetic field \cite{Weyl1922}.} However, there are good reasons for not giving in to this temptation. First of all we have to note that this field couples to the gravitational field in a non-standard way. The vector potential in this theory is thoroughly intertwined with the gravitational field through its presence in the connection and the curvature tensor. Secondly, since any coupling to matter fields has to be done so that conformal invariance is preserved, the vector potential $A_k$ will couple universally to other fields (with the exception of fields which are already conformally covariant, e.g. the electromagnetic field). Thirdly, the electromagnetic field arise normally from gauging a global $U(1)$ symmetry. However, our connection is not a $U(1)$ connection. The Lie group manifold of $U(1)$ has the topology of a circle $S^1$ where $\theta$ and $\theta+2\pi$ are identified. However, such an identification cannot be maintained within the present theory. Indeed, $e^{\theta}g_{ij}\neq e^{\theta+2\pi}g_{ij}$ since the complex unit $i$ is absent in the exponential. Thus the Lie group manifold is not $S^1$ but $\mathbb{R}$. These key mathematical differences indicate that we are not dealing with the electromagnetic field here but rather a new type of universal field.

A fourth reason, for believing that the vector potential must play a different role than the electromagnetic field, comes from cosmology. According to observations the universe is continuously expanding. At first this seems to contradict the very idea of scale invariance: if scale is nothing but an unphysical and unobservable gauge degree of freedom how is it possible that we can determine from observations that the size of the universe is changing? This difficulty is only apparent. There is a fully scale invariant and relational way to understand the expansion of the universe. An apparent expansion of the whole universe can also be understood as the galaxies shrinking in size {\em relative} to the Hubble radius. This ratio of galaxy sizes and the Hubble radius is a scale invariant quantity and does make sense within a scale invariant theory. 

Thus, if our theory can explain the expansion of the universe at all, it must be the case that this vector potential $A_k$ acts effectively as a short-range universal ``shrinking'' force. By ``short-range'' we mean that the shrinking effect should be more pronounced on galactic scales (i.e. near mass concentrations) as compared to the Hubble scale. By ``universal'' we mean that the force should act on all material objects in the same way irrespective of internal composition. Only a universal shrinking force would create the appearance of an expanding universe. As noted above, the vector field has indeed this universal character. However, the electromagnetic field is not a universal field in this sense since it would act on a positively charge body in a different way than on a negatively charged one, and not at all in the case of a neutral body. Thus, it seems clear that the vector potential should not be identified with the electromagnetic field.\footnote{We also believe that the identification of $A_\mu$ with the electromagnetic field within Weyl's theory is also untenable for the same reasons.}

A more likely scenario is for the vector potential to play the role of ``dark matter''. Indeed, there are some indications that scale invariance is important for explaining the galaxy rotation curves \cite{monda, mondb}.\footnote{I'm grateful for Sean Gryb to point this out to me.} It is also of key interest that the spectrum for the microwave background is scale invariant for large wavelengths, i.e. at scales where the mass density is roughly homogenous. Normally, this is explained by an inflationary type cosmological model but in the scale invariant theory presented here it seems natural to expect the spectrum to be scale invariant at scales at which the mass density of the universe is approximately homogenuous. These issues should be explored further.

Furthermore, if the connection $A_k$ is curved so that $\partial_iA_j-\partial_jA_i=F_{ij}\neq0$ then non-trivial global effects would be present. For example, consider two identical rulers at the same point in space. Then move one of them around a closed spatial loop. If we have ``scale curvature'' $F_{ij}\neq0$, then the two rulers might be of different size after this operation. In addition, one would also expect a shift in the spectral lines of hydrogen as discussed in section \ref{hydrogen}. From experience we know that these effects must be very small on human scales but it remains a possibility that they would be observable on cosmological scales. As mentioned in section \ref{hydrogen}, one could speculate that the the tiny increase in the fine structure constant (a controversial claim) could be reinterpreted as a Weyl-type rescaling effect.

Finally we stress again that it is necessary that the notion of proper time can be recovered in some way from the theory. This would presumably happen is if the Hamiltonian constraint could be made to propagate and that the constraints close according to the Dirac-Teitelboim algebra. This would mean that the theory is foliation invariant would probably ensure the emergence of the universal lightcone structure. We could then presumably define a notion of proper time by aid of lightclocks (as sketched in section \ref{proptime}) and in this way we could potentially provide a satisfactory answer to Einstein's objection regarding proper time.
\vspace{0.5cm}
\begin{flushleft}
{\bf \large Acknowledgements}
\end{flushleft}
 This work was supported by FQXI and Perimeter Institute. I have benefitted greatly from discussions with Julian Barbour, Sean Gryb, Lucien Hardy, Tim Koslowksi, Sebastiano Sonego, Rafael Sorkin, and Tom Zlosnik.


\begin{thebibliography}{99}
\bibitem{BB82} J. B. Barbour and B. Bertotti, Proceedings of the Royal Society of London. Series A, Mathematical and Physical Sciences, Vol. {\bf 382}, 295-306, (1982).
\bibitem{BarbourRWR} J. Barbour, ``Relativity without relativity'', Class.Quant.Grav. {\bf 19}, 3217-3248, (2002). arXiv:gr-qc/0012089.
\bibitem{Barbour02partmod} J. Barbour, ``Scale-Invariant Gravity: Particle Dynamics'', Class. Quant. Grav. {\bf 20}, 1543-1570, (2003). gr-qc/0211021. 
\bibitem{Barbour04a} E. Anderson, J. Barbour, B. Foster, B. Kelleher, and N. \'O Murchadha, ``The physical gravitational degrees of freedom'', Class. Quant. Grav. {\bf 22}  1795-1802, (2005). gr-qc/0407104.
 \bibitem{Barbour04b} E. Anderson, J. Barbour, B. Foster, B. Kelleher, N. O'Murchadha, ``A first-principles derivation of York scaling and the Lichnerowicz-York equation''. gr-qc/0404099.
\bibitem{Yorka} J.York, Phys. Rev. Lett. {\bf 26}, 1656, (1971).
\bibitem{Yorkb} J. York, Phys. Rev. Lett. {\bf 28}, 1082, (1972).
\bibitem{Yorkc} J. York, J. Math. Phys. {\bf 14}, 456, (1973).
\bibitem{Anderson} E. Anderson and J. Barbour, ``Interacting vector fields in Relativity without Relativity'', Class. Quant. Grav. {\bf 19},  3249-3262, (2002). gr-qc/0201092.
\bibitem{BSW62} R. Baierlein, D. Sharp, and J. Wheeler, ``Three-Dimensional Geometry as Carrier of Information about time'', Physical Review, Vol. {\bf 126}, No. 5, 1864-5, (1962).
\bibitem{Kiefer} C. Kiefer, ``Quantum Gravity'', International Series of Monographs on Physics 136, 2nd ed. Oxford Science Publications,  (2007).
\bibitem{SonegoWestman} H. Westman, S. Sonego, ``Coordinates, observables and symmetry in relativity'',  Annals of Physics {\bf 324}, 1585-1611,  (2009). arXiv:0711.2651v2 [gr-qc].
\bibitem{Poincare} H. Poincar\'e, ``Science and Hypothesis'', Dover abridged edition (1952).
\bibitem{Weyl1918} H. Weyl, ``Reine Infinitesimalgeometrie'', Math. Z., {\bf 2}, 384-411, (1918).
\bibitem{Weyl1922} H. Weyl, ``Space-Time-Matter'', Dover Publications, fourth edition, (1922).
\bibitem{Bach1921} R. Bach, ``Zur Weylschen Relativitätstheorie und der Weylschen Erweiterung des Krümmungsbegriffs'', Math. Z., {\bf 9}, 110-135, (1921).
\bibitem{Livingrevs} H. Goenner, "On the History of Unified Field Theories", Living Rev. Relativity 7,  (2004),  2. http://www.livingreviews.org/lrr-2004-2 
\bibitem{pagini87} E. Pagini, G. Techhiolli, S. Zerbini, ``On the Problem of Stability for Higher-Order Derivative Lagrangian System'', Lett. in Math. Phys. {\bf 14}, 311--319, (1987). 
\bibitem{Mannheim} P. Mannheim, ``Four Dimensional Conformal Gravity, Confinement, and Galactic Rotation Curves''. gr-qc/9407010.
\bibitem{monda} R. Sanders and S. McGaugh, ``Modified Newtonian Dynamics as an Alternative to Dark Matter'', Ann.Rev.Astron.Astrophys. {\bf 40}  263-317, (2002). astro-ph/0204521.
\bibitem{mondb} M. Milgrom, ``MOND: time for a change of mind?'', Physica Plus 12, arXiv:0908.3842v1 [astro-ph].\\ See also http://physicaplus.org.il/zope/home/en/1223030912/mond\_en (2009). 
\bibitem{Wald} R. Wald, ``General Relativity'', Chicago University Press, (1984). 
\bibitem{Poisson} E. Poisson, ``A Relativists Toolkit'', Cambridge University Press, (2004).
\bibitem{Teitelboim73} C. Teitelboim, ``How commutators of constraints reflect the spacetime structure'', Ann. Phys, {\bf 79}, 542--557, (1973).
\end{thebibliography}
\end{document}